\documentclass[aps,pra,twocolumn,showpacs,groupedaddress,superscriptaddress]{revtex4}
\usepackage{graphicx,graphics,times,amsfonts}
\begin{document}
\title{ Transition of D- Level Quantum Systems Through Quantum Channels with Correlated Noise }
\author{A. Fahmi }
\email{ fahmi@theory.ipm.ac.ir} \affiliation{ Institute for
Studies in Theoretical Physics and Mathematics (IPM) P. O. Box
19395-5531, Tehran, Iran}
\author{M. Golshani }
\email{ golshani@ihcs.ac.ir}  \affiliation{ Institute for Studies
in Theoretical Physics and Mathematics (IPM) P. O. Box 19395-5531,
Tehran, Iran}\affiliation{ Department of Physics, Sharif
University of Technology, P. O. Box 11365-9161, Tehran, Iran}

\begin{abstract}
Entanglement and entanglement-assisted are useful resources to
enhance the mutual information of the Pauli channels, when the
noise on consecutive uses of the channel has some partial
correlations. In this Paper, we study quantum-communication
channels in $d$-dimensional systems and derive the mutual
information of the quantum channels for maximally entangled states
and product states coding with correlated noise. Then, we compare
fidelity between these states. Our results show that there exists
a certain fidelity memory threshold which depends on the dimension
of the Hilbert space $(d)$ and the properties of noisy channels.
We calculate the classical capacity of a particular correlated
noisy channel and show that in order to achieve Holevo limit, we
must use $d$ particles with $d$ degrees of freedom. Our results
show that entanglement is a useful means to enhance the mutual
information. We choose a special non-maximally entangled state and
show that in the quasi-classical depolarizing and quantum
depolarizing channels, maximum classical capacity in the higher
memory channels is given by the maximally entangled state. Hence,
our results show that for high error channels in every degree of
memory, maximally entangled states have better mutual information.
\end{abstract}
 \maketitle
\section{INTRODUCTION}

One of the remarkable by-products of the development of quantum
mechanics in recent years is quantum information and quantum
computation theories. Classical and quantum information theories
have some basic differences. Among these differences are
superposition principle, uncertainty principle, and non-local
effects. The non-locality associated with the entanglement in
quantum mechanics is one of the most subtle and intriguing
phenomena in nature \cite{nil}. Its potential usefulness has been
demonstrated in a variety of applications, such as quantum
teleportation, quantum cryptography, and quantum dense coding
\cite{nil}. On the other hand, quantum entanglement is a fragile
feature, which can be destroyed by interaction with the
environment. This effect, which is due to decoherence \cite{Shor},
is the main obstacle to the practical implementations of quantum
computing and quantum communication. Several strategies have been
devised against decoherence. Quantum error correction codes,
fault-tolerant quantum computation \cite{nil}, and decoherence
free subspaces \cite{Knill} are among them. One of the main
problems in the quantum communication is the decoherence effects
in the quantum channels.

Recently, the study of quantum channels has attracted a lot of
attention. Early works in this direction were devoted, mainly, to
memoryless channels for which consecutive signal transmissions
through the channel are not correlated. The capacities of some of
these channels were determined \cite{Sch,Caves} and it was proved
that in most cases their capacities are additive for single uses
of the channel. For Gaussian channels under Gaussian inputs, the
multiplicativity of the output purities was proved in \cite{Ser}
and the additivity of the energy-constrained capacity, even in the
presence of classical noise and thermal noise, was proved in
\cite{Hir}, under the assumption that successive uses of the
channel are represented by the tensor product of the operators
representing a single use of the channel, i.e., the channel is
memoryless. In a recent letter, Bartlett \emph{et al.} \cite{Bar}
showed that it is possible to communicate with perfect fidelity,
and without a shared reference frame, at a rate that
asymptotically approaches one encoded qubit per transmitted qubit.
They proposed a method to encode a qubit, using photons in a
decoherence-free subspace of the collective noise model. Boileau
\emph{et al.} considered collective-noise channel effects in the
quantum key distribution \cite{Boi1} and they gave a realistic
robust scheme for quantum communication, with polarized entangled
photon pairs \cite{Boi2}. In the last few years much attention has
been given to bosonic quantum channels \cite{Gio}.

Recently, Macchiavello \emph{et al.} \cite{Mac,Bowen}, considered
a different class of channels, in which correlated noise acts on
consecutive uses of channels. They showed that higher mutual
information can be achieved above a certain memory threshold, by
entangling two consecutive uses of the channel. These types of
channels and their extension to the bosonic case, has attracted a
lot of attention in recent years \cite{Gio2}. K. Banaszek \emph{et
al.} \cite{Ball} implemented the suggestion of Macchiavello
\emph{et al.} experimentally. They showed how entanglement could
be used to enhance classical communication in a noisy channel. In
their setting, the introduction of entanglement between two
photons is required in order to maximize the amount of information
that can be encoded in their joint polarization degree of freedom,
and they obtained experimental classical capacity with entangled
states and showed that it is more than $2.5$ times the theoretical
upper limit, when no quantum correlations are allowed. Hence,
recently some people have shown that provided the sender and
receiver share prior entanglement, a higher amount of classical
information is transmitted over Pauli channels in the presence of
memory, as compared to the product and the entangled state coding
\cite{AT}.

On the other hand, there exists a steadily growing interest in
entanglement in higher dimensions, since it allows for the
realization of new types of quantum communication protocols
\cite{Bru}. Those provide more security in quantum communication
\cite{Bec} and quantum key distribution for $d$-level systems,
against individual attacks, in the sense that a slightly higher
error rate is acceptable \cite{Cerf}. Recently, some setups have
been realized by using orbital angular momentum states of photons
\cite{Mair}. Also, transmitting states that belonging to finite
dimensional Hilbert space through quantum channels associated with
a larger Hilbert space \cite{Mancini} and channel capacity in the
higher dimensional systems \cite{Ruskai} have taken a lot of
attention. These considerations encourage us to study various
aspects of quantum information theories at higher dimensions,
e.g., quantum coding, quantum superdense coding, and quantum key
distribution at higher dimensions, through a correlated noisy
channel.

In this paper, we consider the effect of Pauli channels with
correlated noise on the $d$-dimensional systems. We compare
$d$-dimensional maximally entangled and product states with each
other and find a certain memory threshold which depends on the
dimension of Hilbert space $(d)$ and the properties of the noisy
channel. For such states, our results show that in order to reach
a higher fidelity between input and output states, we must use $d$
particles in the $d$-dimensional systems, and fidelity memory
threshold $(\mu^{f}_{t})$ goes to zero for higher dimensional
systems. In the following, we calculate the explicit form of
mutual information of a particular correlated noisy channel and
show that in order to achieve Holevo limit we must make use of $d$
particles with $d$ degree of freedom. Then, entanglement is a
useful means to enhance the mutual information. We choose a
special non-maximally entangled state and show that in the quantum
depolarizing, quasi-classical depolarizing and very high error
channels, maximum classical capacity in the higher memory channels
is given by a maximally entangled state. Our results show that for
high error channels, maximally entangled states have better mutual
information in every degree of memory.

This paper is organized as follows: In Sec. II we briefly review
some properties of quantum channels with correlated noise in the
two dimensional systems. In Sec. III, we extend quantum
communication with correlated noise to $d$-dimensional systems and
calculate fidelity between the input and output states in the
Pauli channels. In Sec. IV, we calculate the mutual information
for the maximally entangled and the product states, for the Pauli
channel of systems with $d$-degree of freedom. In Sec. V, we
discuss optimal properties of quasi-classical depolarizing,
quantum depolarizing and very high error channels. In Sec. VI, we
discuss some applications of quantum channels with correlated
noise in the $d$-dimensional systems. In the remaining part of
this manuscript (Appendixes A,B and C) we derive some useful
relations.

\section{Entanglement-enhanced information transmission over a quantum channel with correlated noise}
Encoding classical information into quantum states of physical
systems gives a physical implementation of the constructs of
information theory. The majority of research into quantum
communication channels has focused on the memoryless case,
although there have been a number of important results obtained
for quantum channels with correlated noise operators or more
general quantum channels \cite{Mac,Ha}. The action of transmission
channels is described by Kraus operators $A_{i}$ \cite{Kraus},
which satisfy the $\sum_{i}A^{\dagger}_{i}A_{i}\leq I$, the
equality holds when the map is trace-preserving. An interesting
class of Kraus operators acting on the individual qubits  can be
expressed in terms of the Pauli operators $\sigma_{x,y,z}$
\begin{eqnarray*}
A_{i} = \sqrt{p_{i}} \sigma_{i}\ ,
\end{eqnarray*}
with $\sum_i p_i = 1$ , $i=0,x,y,z$, and $ \sigma_{0}=I$. A noise
model for these actions is, for instance, the application of a
random rotation by angle $\pi$ around the axis $
\hat{\bf{x}},\hat{\bf{y}},\hat{\bf{z}}$ with the probabilities
$p_x , p_y , p_z$, respectively, and the identity operation with
probability $p_0$. If we send an arbitrary signal $\rho$,
consisting of $n$ qubits (including the entangled ones) through
the channel, the corresponding output state is given by:
\begin{equation}
\rho \longrightarrow \mathcal E(\rho)=\sum_{i_1\cdots i_n}
(A_{i_n}\otimes \cdots \otimes A_{i_1}) \rho
(A^{\dagger}_{i_1}\otimes \cdots \otimes A^{\dagger}_{i_n}).
\end{equation}
In the case of Pauli channels a more general situation is
described by action operators of the following form:
\begin{equation}
A_{k_1\dots k_n} = \sqrt{p_{k_1\dots k_n}} \sigma_{k_1} \dots
\sigma_{k_n}\;,
\end{equation}
with $\sum_{k_1\dots k_n} p_{k_1\dots k_n} =1$. The quantity $
p_{k_1\dots k_n}$ can be interpreted as the probability that a
given random sequence of rotations by angle $\pi$ along the axis
$k_1\dots k_n$ is applied to the sequence of $n$ qubits, sent
through the channel. For a memoryless channel  $p_{k_1\dots k_n}=
p_{k_1}p_{k_2}\dots p_{k_n}$. An interesting generalization is
described by a Markov chain defined as:
\begin{equation}
 p_{k_1\dots k_n} =  p_{k_1} p_{k_2|k_1}\dots  p_{k_n|k_{n-1}}
\end{equation}
where $p_{k_n|k_{n-1}}$ can be interpreted as the conditional
probability that a $\pi$ rotation around the axis $k_n$ is applied
to the $n$-th qubit, given that a $\pi$ rotation around axis
$k_{n-1}$ was applied on the  $n-1$-th qubit. Here we consider the
case of two consecutive uses of a channel with partial memory,
i.e., we shall assume $ p_{k_n|k_{n-1}} =   (1-\mu ) p_{k_n} + \mu
\delta_{k_n|k_{n-1}}$. This means that with probability $ \mu $
the same rotation is applied to both qubits, while with
probability $1 - \mu $ the two rotations are uncorrelated. In the
Macchiavello and co-workers' noise model \cite{Mac}, the degree of
memory $\mu$ could depend on the time laps between the two channel
uses. If two qubits are sent within a very short time interval,
the properties of the channel, which determine the direction of
the random rotations, will be unchanged, and it is therefore
reasonable to assume that the action on both qubits will take the
form:
\begin{equation}
A^{c}_{k} = \sqrt{p_{k}} \sigma_{k} \sigma_{k}.
\end{equation}
If on the other hand, the time interval between the channel uses
is such that the channel properties are changed, then the actions
will be:
\begin{equation}
A^{u}_{k_1,k_2} = \sqrt{p_{k_1}}\sqrt{p_{k_2}}
\sigma_{k_1}\sigma_{k_2}.
\end{equation}
An intermediate case, as mentioned above, is described by actions
of the form:
\begin{equation}
A^{i}_{k_{n-1},k_{n}} = \sqrt{p_{k_{n-1}} [(1-\mu ) p_{k_{n}} +
\mu \delta_{k_n|k_{n-1}} ] } \sigma_{k_{n-1}}\sigma_{k_{n}}.
\label{part}
\end{equation}
They have shown that the transmission of classical information can
be enhanced by employing maximally entangled states as carriers of
information, rather than the product states. Hence, they obtained
a threshold in the degree of memory above which a higher amount of
classical information is transmitted with entangled signals
\cite{Mac}.

\section{Fidelity Between $D$-Dimensional Input and Output States in the Pauli Channels with Correlated Noise}

In this section, we would like to compute the fidelity between
input and output of $d$-dimensional systems with correlated noise
for maximally entangled states and product states. At first, it
seems that for generalization of the previous works \cite{Mac} to
$d$-dimensional systems, we must use two particles with
$d$-degrees of freedom  \cite{Cerf1}, but as we shall see in the
next section, it is more appropriate to use $d$ particles each
having $d$ degrees of freedom.

Let us consider first the effect of the Pauli channels with
correlated noise on $d$-particles in $d$-dimensional states.
Maximally entangled states $|\psi_{[l_{i},s]}\rangle$ are a set of
$(d^{d})$ orthonormal Hilbert spaces of $d$ qudits:
\begin{eqnarray}\label{Maxim}
&&|\psi_{[l_{i},s]}\rangle=\frac{1}{\sqrt{d}}\sum_{j=0}^{d-1} e^{2
\pi i \frac{ js}{d}}|j\rangle
|j+l_{1}\rangle...|j+l_{d-1}\rangle\nonumber\\
&&\langle\psi_{[l'_{i},s']}|\psi_{[l_{i},s]}\rangle=\delta_{s,s'}\prod_{i=1}^{d-1}\delta_{l_{i},l'_{i}}
\end{eqnarray}
where $[l_{i},s]$ are used for summarizing the representation of
the states with $l_{i},s= 0,..., d-1$, $i=1,...,d-1$, and kets
must be taken \emph{modulo} $d$ here. Quantum channel effects on
the qudits are represented by $U_{m,n}$, generalizing the Pauli
matrices for qubits to $d$-dimensional systems, which are a group
of qubit error operators defined as:
\begin{eqnarray}\label{U}
U_{m,n} = \sum_{k=0}^{d-1} e^{2 \pi i \frac{ kn}{d}}|k+m
\rangle\langle k|.
\end{eqnarray}
Here, $m$ labels the shift errors (extending the bit flip
$\sigma_{x}$) and $n$ labels the phase errors (extending the phase
flip $\sigma_{z}$). We could consider a simple set of error
operators which are used in multidimensional quantum
error-correction codes \cite{Gott}. In this approach a mixture of
pure states can give the same results \cite{Fa}. The effect of the
coupling between the qudits and environment can be absorbed in a
dielectric coefficient (for example, photons of atmosphere or
optical fiber). The dielectric constant has a spatial and temporal
dependence, leading to an overall time dependent unitary
transformation of the polarization state of a single qudit
$A(t)_{m,n}$, as the net effect of the channel. The effect of this
noise on the state of qudits can be considered by:
\begin{eqnarray*}
\rho_{[l_{j},s]} \longrightarrow \mathcal
E(\rho_{[l_{j},s]})=\sum_{[m_{i},n_{i}]}A^{\mu}_{[m_{i},n_{i}]}
\rho_{[l_{j},s]} A^{\mu \hspace{.1cm} \dagger }_{[m_{i},n_{i}]}.
\end{eqnarray*}
Here $ \rho_{[l_{j},s]}=|\psi_{[l_{j},s]}\rangle
\langle\psi_{[l_{j},s]}| $ and $ A^{\mu}_{[m_{i},n_{i}]}$ are
extending Kraus operators, which satisfy the relation
$\sum_{[m_{i},n_{i}]}A^{\mu\hspace{.1cm}
\dagger}_{[m_{i},n_{i}]}A^{\mu}_{[m_{i},n_{i}]}=I$, where $[m_{i},
n_{i}]$ with $i= 0,...,d-1$ and $m_{i}, n_{i}= 0,...,d-1$,
summarize the representation of all particles.

We consider the shortest variational time of the channel (fiber)
under thermal and mechanical fluctuations as $\tau_{fluc}$. If the
time lapse ($t_{lap}$) between the two channels used is small
compared to $\tau_{fluc}$, the effects of the channel on various
qudits can be considered with correlated noise. For example, in
the experiment of K. Banaszek \emph{et al.} \cite{Ball}
$t_{lap}\approx 6\hspace{1mm} ns$, which is much smaller than the
mechanical fluctuations of the fiber. For simplicity, we consider
only two types of sending particles: (I) $d$ particles in Alice's
hand are sent at the same time ($\tau_{lap}\ll\tau_{fluc}$ for
each pair); (II) $d$ particles are sent with a time delay
($\tau_{fluc}\ll \tau_{lap}$ for each pair). Although we can
consider the general case of the transmission of quantum states,
its calculation is very complicated and does not clarify any
physical properties. Similar to the previous cases, Karus
operators $A^{\mu}_{[m_{i},n_{i}]}$ in the presence of partial
memory are described by:
\begin{widetext}
\begin{eqnarray}\label{Kar}
A^{\mu}_{[m_{i},n_{i}]}=
\left[(1-\mu)\prod_{i=0}^{d-1}P_{m_{i},n_{i}} + \mu
P_{m_{0},n_{0}}\prod_{i=0}^{d-1} \delta_{m_{i},m_{i+1}}
\delta_{n_{i},n_{i+1}} \right]^{\frac{1}{2}}
U_{m_{0},n_{0}}...U_{m_{d-1},n_{d-1}}
\end{eqnarray}
\end{widetext}
where $P_{m_{i},n_{i}}$ can be interpreted as the probability that
error $U_{m_{i},n_{i}}$ is applied on the $i-$th qudit, and $\mu$
could be considered as the degree of memory of the the channel,
which can depend on the time lapse between the two channels used.
For the early realization of channels with correlated noise, we
would like to compare the similarity between the input and the
output states. Fidelity can be considered as a measure for
comparison. It will be shown that above a certain memory threshold
$\mu^{f}_{t}$, we have higher fidelity for maximally entangled
states.

The fidelity between the input and the output states can be
expressed as: $F_{[l_{i},s]} = \langle\psi_{[l_{i},s]}|\mathcal
E(\rho_{[l_{i},s]})|\psi_{[l_{i},s]}\rangle$, which for maximally
entangled states (\ref{Maxim}) becomes:
\begin{eqnarray*}
F^{max-en}_{[l_{i},s]} =&&(1-\mu)\frac{1}{d^{2}}
\sum_{j_{1},j_{2},m=0}^{d-1} \left[\sum_{n=0}^{d-1} P_{m,n} e^{
\frac{2\pi i}{d}(j_{1}-j_{2})n} \right]^{d}\nonumber\\&& + \mu.
\end{eqnarray*}
We consider a $(d\times d)$ probability parameter $P_{m,n}$,
represented by the matrix $P$, which is a generalization of the
form given in \cite{Cerf}:
\begin{eqnarray}\label{Mat}
P=\left(%
\begin{array}{cccccc}
  p & q & . & . & . & q \\
  r & t & . & . & . & t \\
  . & . & . & . & . & . \\
  . & . & . & . & . & . \\
  . & . & . & . & . & . \\
  r & t & . & . & . & t \\
\end{array}%
\right)
\end{eqnarray}
where $p$ represents the probability of the transition without any
errors, and $r$ and $q$ represent the probabilities of the shift
and phase errors, respectively, and $t$ represents the probability
of mutual shift and phase errors, where $p$, $q$, $r$, and $t$ are
real variables satisfying the normalization condition
$p+(d-1)q+(d-1)r+(d-1)^{2}t=1$. We can consider $d$-dimensional
unitary operators, which were considered by Gottesman \emph{et
al.} \cite{Gott,Fa1}. The explicit form of
$F^{max-en}_{[l_{i},s]}$ is:
\begin{eqnarray}
&&F^{max-en}_{[l_{j},s]}=\frac{(1-\mu)}{d}(p-q)\nonumber\\
&&\times\left\{(p-q)^{d-1}(d-1)+(p-q+qd)^{d-1}\right\}\nonumber\\&&+
\frac{(1-\mu)}{d}(r-t)(d-1)\nonumber\\&&\times
\left\{(r-t)^{d-1}(d-1)+(r-t+td)^{d-1}\right\}\nonumber\\
&&+\frac{(1-\mu)}{d^{2}}
\left\{td^{2}(p-q+qd)^{d-1}\right.\nonumber\\&&\left.+td(d-1)(r-t+td)^{d-1}(td)^{d}(d-1)^{2}\right\}\nonumber\\&&+(1-\mu)(q-t)
[p-q+qd]^{d-1} + \mu
\end{eqnarray}
For product states, we have some different choices \cite{Com}, if
we consider the computational basis ${|s_{i}\rangle}$ given by:
\begin{eqnarray}\label{Pro}
|\phi^{pro}_{[s_{i}]}\rangle=\bigotimes_{i=0}^{d-1}|s_{i}\rangle
\end{eqnarray}
where $\bigotimes_{i=0}^{d-1}|s_{i}\rangle$ is used for
representing the tensor product of $d$ particles with $d$ degrees
of freedom, with $s_{i}= 0,..., d-1$, and $i=0,...,d-1$ ($d^{d}$
orthonormal product states of $d$ qudits). By using Eq.
(\ref{Mat}), the fidelity for the product states is given by:
\begin{eqnarray}
F^{pro}_{[s_{i}]} = (1-\mu)[p+(d-1)q]^{d} +
\mu[p+(d-1)q]\nonumber\\
\end{eqnarray}
Here $F^{max-en}$ and  $F^{pro}$ do not depend on the $[l_{i},s]$
and $[s_{i}]$. Thus, we can consider the above fidelity for all
maximally entangled and product states bases, respectively. Hence,
$F^{max-en}$ and  $F^{pro}$ are equal to their maximum value
$(F^{max-en}=F^{pro}=1)$ for noiseless channels $(p=1)$. Hence,
for high memory channels $(\mu^{f}=1)$ only $F^{max-en}$ goes to
its maximum value. The explicit form of $\mu^{f}_{t}$ is very
complicated. For a special case of $q=r=t$, the memory threshold
$\mu^{f}_{t}$ for the same fidelity of states (\ref{Maxim}) and
(\ref{Pro}), as a function of $(p,q,d)$, is given by:
\begin{widetext}
\begin{eqnarray*}
\mu^{f}_{t}=\frac{ \frac{d-1}{d}\left\{(p-q+qd)^{d} -
(p-q)^{d}-(qd)^{d}\right\}} {
1-(p-q+qd)+\frac{d-1}{d}\left\{(p-q+qd)^{d} -
(p-q)^{d}-(qd)^{d}\right\}}
\end{eqnarray*}
\end{widetext}
In the above relation, for highly noisy channels $(p=q)$ and for
very low noisy channels $(p\gg q)$ with $d \longrightarrow
\infty$, the fidelity memory threshold goes to zero
$\mu^{f}_{t}\longrightarrow 0$; in other words, similar to the
previous works of \cite{Cerf}, we can show that at higher
dimensions, the use of maximally entangled states are very
suitable for quantum communication.
\section{Holevo Limit for the $D$-Dimensional Pauli Channels with Correlated Noise}
In this section, we would like to compute the amount of
information that can be transmitted through a quantum channel with
correlated noise. We shall derive a general expression for the
mutual information in the quantum channel for the maximally
entangled state (\ref{Maxim}) and the product state (\ref{Pro}) by
using Kraus operators $A^{\mu}_{[m_{i},n_{i}]}$ (\ref{Kar}) and
probability parameters, which are defined by (\ref{Mat}).

The mutual information $I(\mathcal E (\rho_{[l_{i},s]}))$ of a
general quantum channel $\mathcal E$ is given by:
\begin{widetext}
\begin{eqnarray}\label{Hol}
I(\mathcal E (\rho_{[l_{i},s]}), {\pi_{[l_{i},s]}})=
S(\sum_{[l_{i},s]}\pi_{[l_{i},s]}\mathcal E (\rho_{[l_{i},s]}))
-\sum_{[l_{i},s]}\pi_{[l_{i},s]}S(\mathcal E(\rho_{[l_{i},s]})).
\end{eqnarray}
\end{widetext}
Hence, it is the maximum value which is called the
Holevo-Schumacher-Westmoreland channel capacity, and is defined as
\cite{Sch}:
\begin{eqnarray}
\chi(\mathcal E, {\pi_{[l_{i},s]}})=Max_{[ \pi, \rho ]} I(\mathcal
E (\rho_{[l_{i},s]}), {\pi_{[l_{i},s]}})
\end{eqnarray}
where $S(\omega)=-Tr(\omega\log_{2}\omega)$ is the von Neumann
entropy of the density operator $\omega$ and the maximization is
performed over all input probability distribution
$\pi_{[l_{i},s]}$ and density matrix $\rho_{[l_{i},s]}$. Note that
this bound incorporates maximization over all positive operator
value measures (POVM) measurements at the receiver, including the
collective ones over multiple uses of the channel. In what
follows, we shall derive $I(\mathcal E (\rho_{[l_{i},s]}))$ for
maximally entangled and product states. We know that the maximally
entangled input states $|\psi^{max-en}_{[l_{j},s]}\rangle$ can be
derived from $|\psi^{max-en}_{[0,0]}\rangle$, by unitary
transformations, and with respect to the von Neumann entropy,
$S(\omega)$ is invariant under any unitary transformation of a
quantum state $\omega$. The second term on the right hand side of
Eq.(\ref{Hol}), for the states which are related to each other by
a unitary transformation $U_{m,n}$ (especially maximally entangled
states) becomes:
\begin{eqnarray*}
\sum_{[l_{i},s]}\pi_{[l_{i},s]}S(\mathcal E(\rho_{[l_{i},s]}))=
S(\mathcal E(\rho_{[0,0]})).
\end{eqnarray*}
On the other hand, with respect to the relation $Tr_{[l_{i},s,
i\neq j]}\rho_{[l_{i},s]}=\frac{1}{d}\textbf{I}_{j}$, we show that
the Holevo limit can be attained by setting
$\pi_{[l_{i},s]}=\frac{1}{d^{d}}$ (with $l_{i},s=0,...,d-1$). The
quantum state $\mathcal E(\rho_{[l_{i},s]})$ in the first term can
be written as \cite{Ban}:
\begin{eqnarray*}
\sum_{[l_{i},s]}\pi_{[l_{i},s]}\mathcal
E(\rho_{[l_{i},s]})&&\nonumber\\
=\frac{1}{d^{d}}\sum_{[l_{i},s]}\mathcal E(\rho_{[l_{i},s]})
&=&\mathcal
E(\frac({1}{d}\textbf{I}_{0}\otimes...\otimes\frac{1}{d}\textbf{I}_{d-1})
\end{eqnarray*}
To get the final result, we take suitable bases for density matrix
representation.

If we consider maximally entangled states as input states Eq.
(\ref{Maxim}) for $s,l_{i}=0$ (with $i=1,...,d-1$), which is
represented by $|\psi^{max-en}_{[0,0]}\rangle$, for channels with
correlated noise, the output states are given by:
\begin{widetext}
\begin{eqnarray}\label{Outen}
\mathcal
E^{max-en}(\rho_{[0,0]})&=&\frac{(1-\mu)}{d}\sum_{j_{1},j_{2}=0}^{d-1}
\left[\sum_{m_{0},n_{0}=0}^{d-1}e^{\frac{2\pi
i}{d}(j_{1}-j_{2})n_{0}}P_{m_{0},n_{0}}|j_{1}+m_{0}\rangle\langle
j_{2}+m_{0}|\right]^{\otimes d}+\mu\rho_{[0,0]}\nonumber\\
&=&\frac{(1-\mu)}{d}\sum_{j_{1},j_{2}=0}^{d-1}
\left[(p-q)|j_{1}\rangle\langle
j_{2}|+(r-t)\sum_{m=1}^{d-1}|j_{1}+m\rangle\langle j_{2}+m|
+(q-t) \delta_{j_{1},j_{2}}|j_{1}\rangle\langle
j_{2}|+td\delta_{j_{1},j_{2}}I \right]^{\otimes d}\nonumber\\
&&+\mu\frac{1}{d}\sum_{j_{1},j_{2}=0}^{d-1} |j_{1}\rangle\langle
j_{2}|^{\otimes d}.
\end{eqnarray}
\end{widetext}
In the above relation, $[\sigma]^{\otimes k}$ represents the
tensor product of $k$,$\sigma$ matrices and we have used
probability parameters suggested by Eq. (\ref{Mat}). Furthermore,
we would like to calculate the entropy function for the above
output density matrix. For simplicity of our calculations, we
apply some unitary transformations (\textsc{c-not} operator) on
the above density matrix (it is known that entropy function does
not change by any unitary transformation). As we have calculated
in Appendix A, after \textsc{c-not} unitary transformations, the
mutual information is given by the following equation:
\begin{eqnarray}
I^{max-en}(\mathcal E (\rho_{[l_{i},s]})) &=& d\log_{2}d+
\sum_{[k_{i}=0]}^{d-1}\Lambda^{0}_{[k_{i}]}\log_{2}\Lambda^{0}_{[k_{i}]}\nonumber\\&+&
(d-1)\sum_{[k_{i}=0]}^{d-1}\Lambda^{1}_{[k_{i}]}\log_{2}\Lambda^{1}_{[k_{i}]}
\end{eqnarray}
In the above relation, $[k_{i}]$ represents a set of variables
with $i=1,...,d-1$, and $\Lambda^{l}_{[k_{i}]}$ given by:
\begin{eqnarray}
\Lambda^{0}_{[k_{i}]}=dB_{[k_{i}]}+A_{[k_{i}]}\hspace{1cm}
\Lambda^{l}_{[k_{i}]}=A_{[k_{i}]},
\hspace{.4cm}l=1,...,d-1\nonumber\\
\end{eqnarray}
$A_{[k_{i}]}$ and $B_{[k_{i}]}$ have been defined in Appendix A.

In a similar manner, if the input state considered is the product
state defined by Eq. (\ref{Pro}), then the output density matrix
is given by:
\begin{widetext}
\begin{eqnarray}
\mathcal
E^{pro}(\varrho_{[0]})&=&(1-\mu)\sum_{[m_{i},n_{i}]}\prod_{i=1}^{d-1}P_{m_{i},n_{i}}|m_{i}\rangle\langle
m_{i}|^{\otimes d}+\mu\sum_{m,n}P_{m,n}|m\rangle\langle
m|^{\otimes d} \\&&
=(1-\mu)\left[\sum_{[m,n]}P_{m,n}|m\rangle\langle
m|\right]^{\otimes d}+\mu\sum_{[m,n]}P_{m,n}|m\rangle\langle
m|^{\otimes d}\nonumber\\
\mathcal
E^{pro}(\varrho_{[0]})&=&(1-\mu)\left[[x+qd]|0\rangle\langle
0|+[y+dt]\sum_{m=1}^{d-1}|m\rangle\langle m|\right ]^{\otimes d}
+\mu\left[[x+qd]|0\rangle\langle 0|^{\otimes
d}+[y+dt]\sum_{m=1}^{d-1}|m\rangle\langle m|^{\otimes
d}\right]\nonumber
\end{eqnarray}
\end{widetext}
with $x=p+(d-1)q$ and $y=qd$, and after some simple algebra, the
mutual information of the above density matrix is given by:
\begin{widetext}
\begin{eqnarray}
I^{pro}(\mathcal E (\varrho_{[s_{i}]})) &=&
d\log_{2}d+\left[(1-\mu)(x+qd)^{d}+\mu
(x+qd)\right]\log_{2}\left[(1-\mu)(x+qd)^{d}+\mu
(x+qd)\right]\nonumber\\&&+(d-1)\left[(1-\mu)(y+dt)^{d}+\mu
(y+dt)\right]\log_{2}\left[(1-\mu)(y+dt)^{d}+\mu
(y+dt)\right]\nonumber\\&&+\sum_{k=0}^{d-1}\left[(d-1)^{d-k}\left(%
\begin{array}{c}
  d \\
  k \\
\end{array}%
\right)-(d-1)\delta_{k,0}\right]\left[(1-\mu)(x+qd)^{k}(y+dt)^{d-k}\right]\nonumber\\&&
\times\log_{2}\left[(1-\mu)(x+qd)^{k}(y+dt)^{d-k}\right].\nonumber\\
\end{eqnarray}
\end{widetext}
In the next section, we shall restrict ourselves to a
quasi-classical depolarizing channel, a quantum depolarizing
channel, and a very high error channel and discuss their
properties.

\section{Discussion on optimization}
In this section, we consider quasi-classical depolarizing channels
and quantum depolarizing channels and suggest a non-maximally
entangled state that interpolates between the product state and
the maximally entangled state and show that the mutual information
is monotonously modified when this state goes from a product state
to a maximally entangled state. With an overview on the output
density matrix of the channel, we see that the mutual information
in the $\mu=0$ case (channels without memory), the product states
in the computational basis \cite{Com} are the most suitable
states; on the other hand, in the $\mu=1$ case (channels with
completely correlated noise), maximally entangled states are the
most suitable states for quantum communications. These two mutual
informations cross over each other at the point $\mu_{t}$. It must
be shown that mutual information for each non-maximally entangled
state does not cross mutual information of the product states for
$\mu<\mu_{t}$ and the maximally entangled state for $\mu>\mu_{t}$.
In the following, we discuss various types of non-maximal input
states and show that they have aforementioned properties and in
the special case of high error channels $p=q=r=t$, we prove the
above properties.

\subsection{Quasi-Classical Depolarizing Channel}
A quasi-classical depolarizing channel is given by the following
probability parameters, which are same as the probability
parameters in Eq. (\ref{Mat}), with $p=q$ and $r=t$:
\begin{equation}\label{}
p_{m,n}= p_m = \left\{\begin{array}{ll}
p=q  &,\quad m=0, \\
{\displaystyle r=t=\frac{1-dp}{d(d-1)}} &,\quad
\textnormal{otherwise} .
\end{array}
\right.
\end{equation}
Similar to two qubit \cite{Mac} and two qudit cases \cite{Cerf1},
we define $\rho_{\ast}$ to denote a chosen input state giving
minimal output entropy, when transmitted through the channel
$\mathcal E$. We consider $U_{m_{i},n_{i}}$ as an irreducible
representation of a compact group with
$U_{m_{i},n_{i}}U_{m_{j},n_{j}}= e^{2\pi i
(m_{j}n_{i}-m_{i}n_{j})/d}U_{m_{j},n_{j}}U_{m_{i},n_{i}}$ and
quantum channels that are covariant with respect to this compact
group. We consider averaging the operator ${\mathcal F}$
\begin{figure}
\centering
\includegraphics[height=7.15cm,width=8cm]{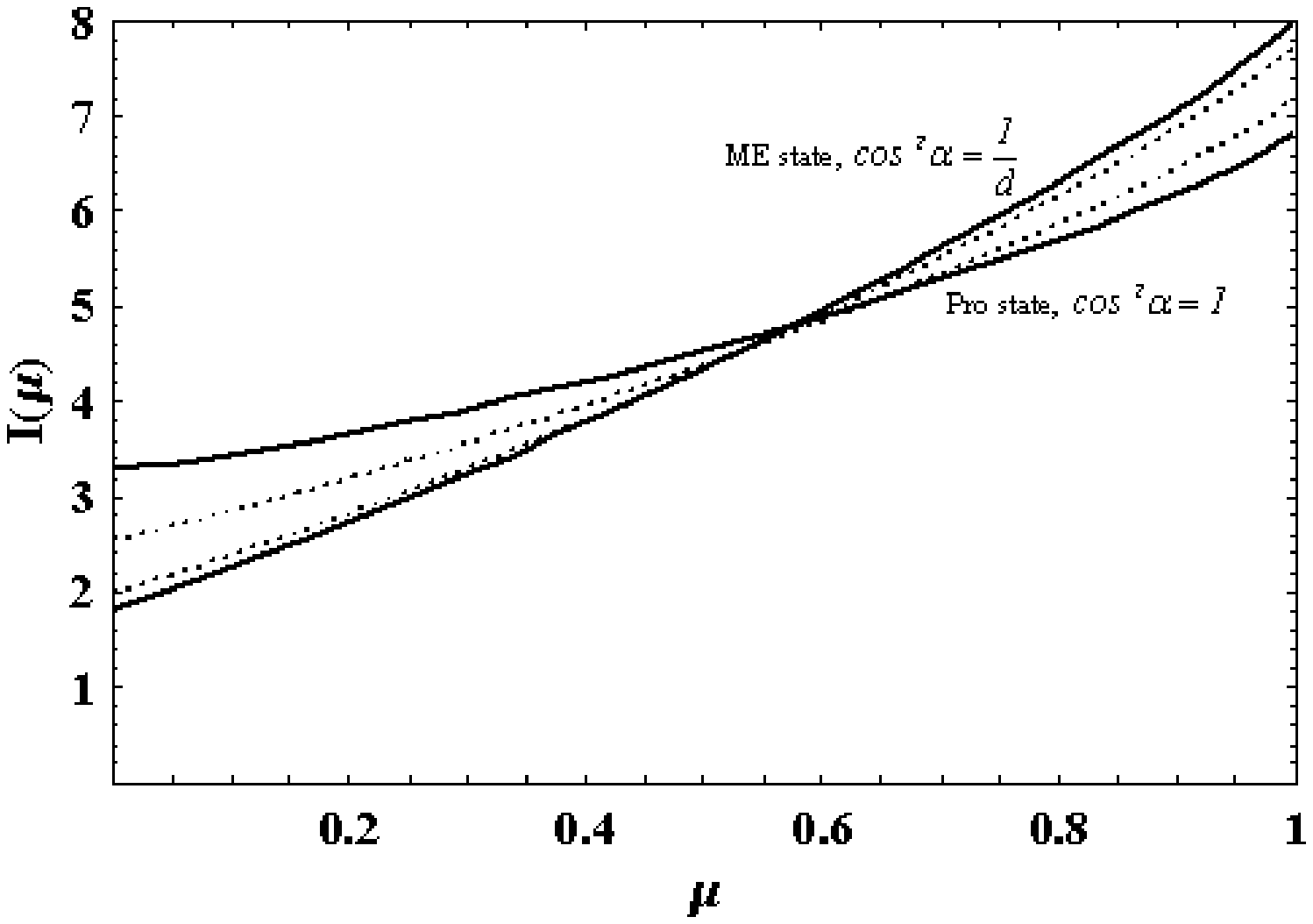}
\caption{Mutual information $I({\mathcal E}_2(\rho_\alpha))$ as a
function of the memory  parameter $\mu$ for the quasi-depolarizing
channel with  $p=0.15$, for different values of the optimization
parameter $\alpha$, in the $d=4$ dimensional systems. Solid lines
represent the mutual information for maximally entangled and
product states and the dashed lines are used for non-maximally
entangled states with $\cos^{2}\alpha=\sqrt{\frac{1}{n}}$,
$n=1.25, 2$.} \label{QDC-D}
\end{figure}
\begin{equation}
{\mathcal F}(\rho)=\frac{1}{d^{d}}\sum_{[n_{i}=0]}^{d-1}
(U_{0,n_{0}}...U_{0,n_{d-1}})\rho(U^{\dagger}_{0,n_{0}}...U^{\dagger}_{0,n_{d-1}}).
\end{equation}
It is not complicated to show that ${\mathcal F}(\rho)$ does not
affect the output of the quantum channel, in the sense that:
\begin{eqnarray*}\label{}
\mathcal E_2\circ{\mathcal F}=\mathcal E_2,
\hspace{0.5cm}S(\mathcal E (\rho_{\ast}))=S(\mathcal
E\circ{\mathcal F(\rho_{\ast})})
\end{eqnarray*}
Thus, if $\rho_{*}$ is an \emph{optimal} state, then ${\mathcal
F}(\rho_*)$ is also an optimal state. Therefore, we can restrict
our search to the whole space ${\mathcal H}^{\otimes d}$ to
${\mathcal F}({\mathcal H}^{\otimes d})$. Finally, using
(\ref{U}), it is straightforward to show that any state  from
${\mathcal F}({\mathcal H}^{\otimes d})$ is a convex combination
of pure states
$|\omega_{[l_{i},s]}\rangle\langle\omega_{[l_{i},s]}|$ where:
\begin{eqnarray}\label{25}
|\omega_{[l_{i},s]}\rangle
=\sum_{j=0}^{d-1}a_{j}e^{i\phi_j}|j\rangle
|j+l_{1}\rangle...|j+l_{d-1}\rangle,\nonumber\\ \quad a_j\in
\mathbb{R}, \quad \sum^{d-1}_{j=0} a_j^2=1 .
\end{eqnarray}
Restricting our search to the states of the form (\ref{25}), we
reduce the number of real optimization parameters from $(2d)^d$ to
$2d$, which can still be a large number. In order to reduce this
number to 1, we consider the following ansatz \cite{Cerf1}:
\begin{equation}\label{SUST}
|\psi(\alpha)\rangle =\cos\alpha|0\rangle^{\otimes d} +
\frac{\sin\alpha}{\sqrt{d-1}} \sum^{d-1}_{j=1}|j\rangle^{\otimes
d}.
\end{equation}
interpolating between the product state ($\cos\alpha=0$) and the
maximally entangled state ($\cos^2\alpha=1/d$). Using the
one-parameter family of input states
$\rho_\alpha=|\psi(\alpha)\rangle\langle\psi(\alpha)|$, in Fig.
(\ref{QDC-D}), we show the mutual information $I(\mathcal
E_2(\rho_\alpha))$ for different values of $\alpha$ (Appendix B).
The mutual information is monotonously modified when $\alpha$ goes
from a product state to a maximally entangled state, whereas the
crossover point $\mu_t$ stays intact. However, we cannot guarantee
that no other configuration of the parameters $a_j$ and $\phi_j$
minimizes the entropy $S(\mathcal E_2(\rho))$, and this provides
the maximum of the mutual information.

\begin{figure}
\centering
\includegraphics[height=7cm,width=8cm]{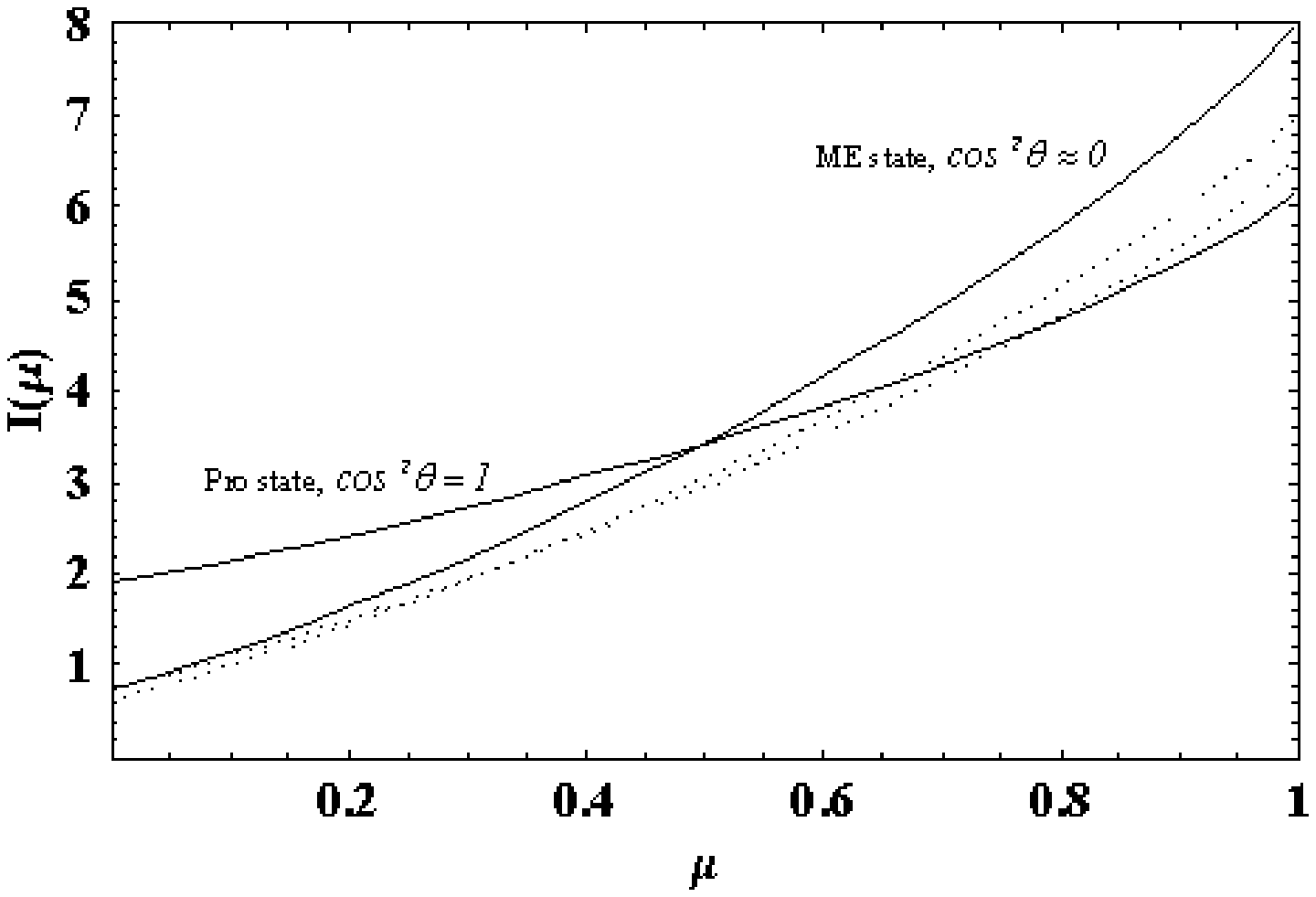}
\caption{Mutual information $I({\mathcal E}_2(\rho_\theta))$ as a
function of the memory  parameter $\mu$ for the quasi-depolarizing
channel with $p=0.2$ for the state of Eq. (\ref{KM1}) and
different values of the optimization parameter $\theta$, in the
$d=4$ dimensional systems. Solid lines represent mutual
information for maximally entangled and product states and the
dashed lines are used for non-maximally entangled states with
$\cos^{2}\theta=\sqrt{\frac{1}{n}}$, $n=2, 4$.} \label{QDC-D=4}
\end{figure}

As another example, we consider other states that continuously
interpolate between the product basis and the maximally entangled
basis (at least in the special case) \cite{KM}. We consider, e.g.,
the single state $|\Psi_{[0,0]}\rangle=\sum_{j=0}^{d-1}
A_{j}|j\rangle^{\otimes d}$, as an input state. With some revision
in the coefficients in the \cite{KM}, for the $d=4$ case (complex
coefficients), these coefficients would be \cite{KMCom}:
\begin{equation}\label{KM1}
A_{0}=\frac{1}{2}(1+e^{i\theta}\cos\theta),
\hspace{0.5cm}A_{1}=-iA_{2}=-A_{3}=\frac{1}{2}e^{i\theta}\sin\theta
\end{equation}
We apply the above coefficients in the output density matrix
(\ref{B1}) and derive the mutual information (its explicit form is
derived in Appendix B). In Fig.(\ref{QDC-D=4}), we plot the mutual
information for various amounts of $\theta$, versus its memory
coefficient $\mu$. Fig.(\ref{QDC-D=4}) shows that the use of
product states for $\mu<\mu_{t}$ and maximally entangled states
for $\mu>\mu_{t}$ are more appropriate for communication,
although, the crossover point $\mu_{t}$ doesn't stay fixed for
various $\theta$.
\subsection{Quantum Depolarizing Channel}
In this subsection, we would like to discuss the depolarizing
channels and to show how to derive an explicit expression for the
depolarizing channel with correlated noise. This channel is given
by the following probability parameters which are the same as in
Eq. (\ref{Mat}), with $q=r=t$:
\begin{equation}\label{}
p_{m,n} = \left\{\begin{array}{ll}
p & , \quad m=n=0 , \\
{\displaystyle q=r=t= \frac{1-p}{d^2-1}} & , \quad
\textnormal{otherwise}.
\end{array}
\right.
\end{equation}
\begin{figure}
\centering
\includegraphics[height=7cm,width=8cm]{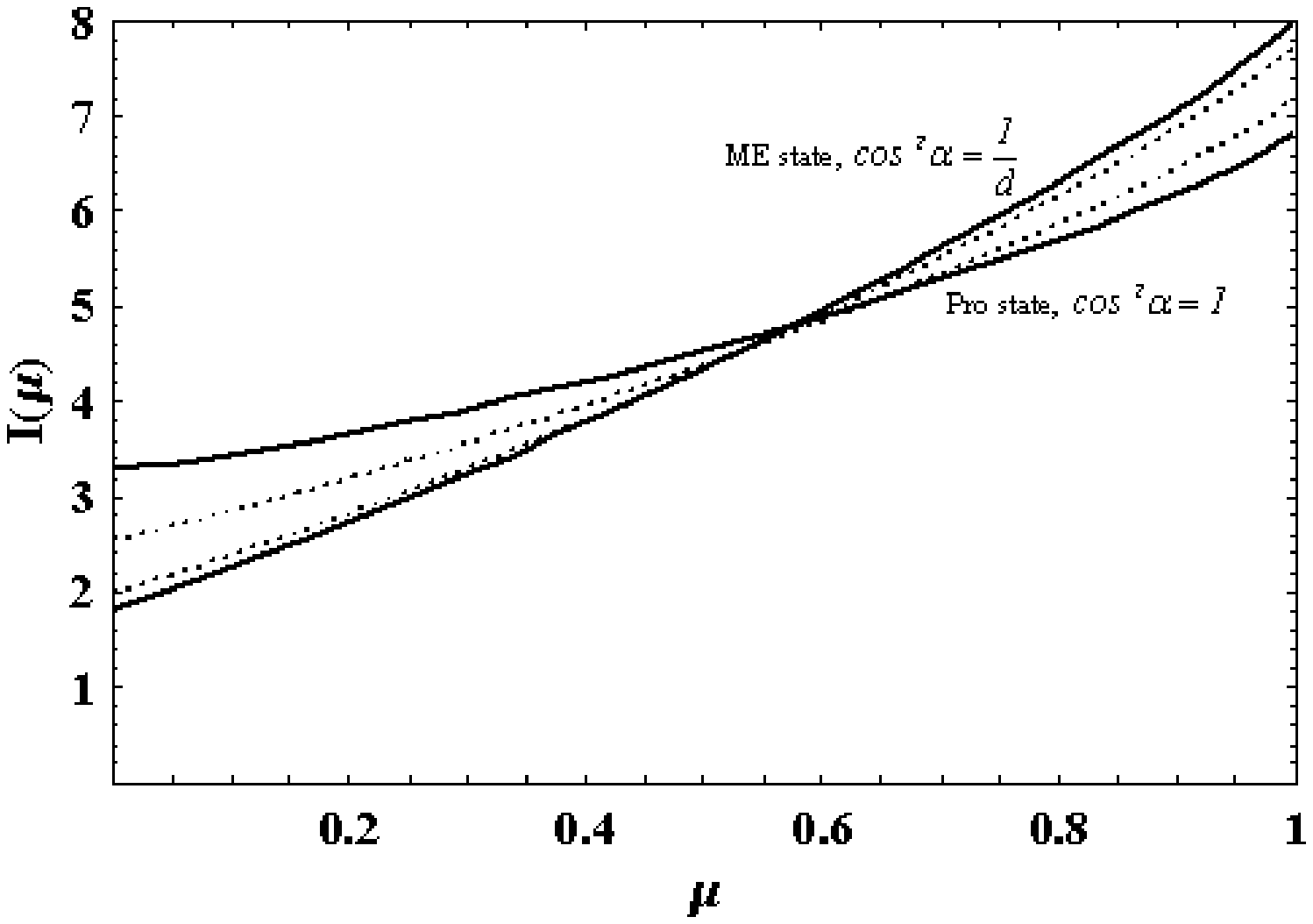}
\caption{Mutual information $I({\mathcal E}_2(\rho_\alpha))$ as a
function of the memory  parameter $\mu$ for the quantum
depolarizing channel with $p=0.7$, for different values of the
optimization parameter $\alpha$, in the $d=4$ dimensional systems.
The solid lines represent the mutual information for maximally
entangled and product states and the dashed lines are used for
non-maximally entangled states with
$\cos^{2}\alpha=\sqrt{\frac{1}{n}}$, $n=1.25, 2$.}\label{DC-D1}
\end{figure}

\begin{figure}
\centering
\includegraphics[height=7cm,width=8cm]{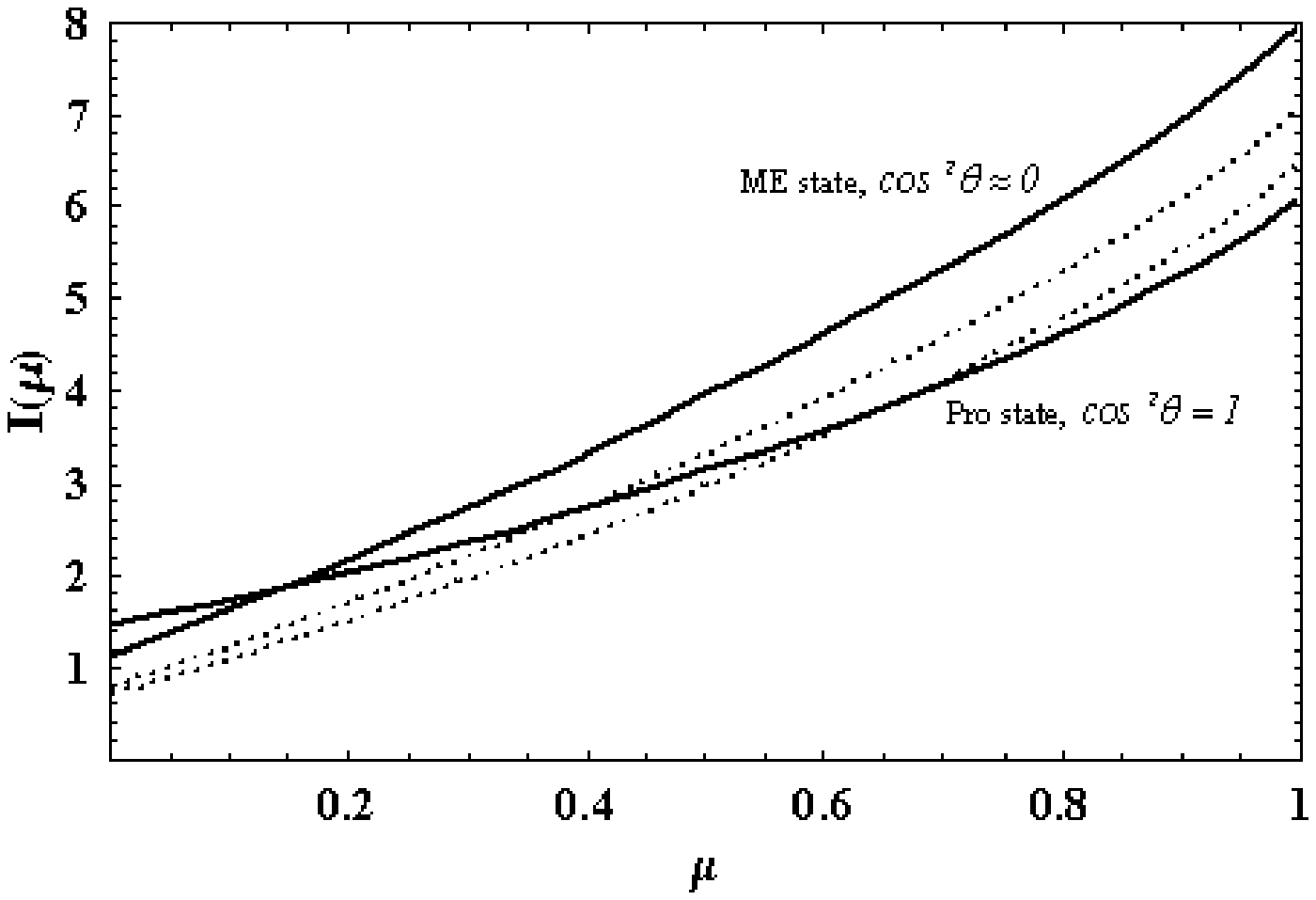}
\caption{Mutual information $I({\mathcal E}_2(\rho_\theta))$ as a
function of the memory  parameter $\mu$ for the quantum
depolarizing channel with $p=0.7$,  for the state of Eq.
(\ref{KM1}) different values of the optimization parameter
$\theta$, in the $d=4$ dimensional systems. The solid lines
represent the mutual information for maximally entangled and
product states and the dashed lines are used for non-maximally
entangled states with $\cos^{2}\theta=\sqrt{\frac{1}{n}}$, $n=2,
5$.}\label{DC-D=4}
\end{figure}

Although a discussion for quantum depolarizing channels similar to
the case of quasi-classical depolarizing channels would be
complicated, but, we consider (\ref{SUST}) as an input state for
quantum depolarizing channels and discuss the output density
matrix and mutual information for various amounts of $\alpha$. The
output density matrix and the explicit form of the mutual
information are derived in the Appendix C. In Fig. (\ref{DC-D1}),
we present the mutual information $I(\mathcal E_2(\rho_\alpha))$
for different values of $\alpha$. Similar to quasi-classical
depolarizing channels, the mutual information is monotonously
modified when $\alpha$ goes from a product state to a maximally
entangled state. In Fig. (\ref{DC-D=4}), we plot the mutual
information of the quantum depolarizing channel for the state
coefficients that were suggested in the Eq. (\ref{KM1}) and an
explicit form of them is derived in Appendix C.

\subsection{Optimal Quantum Communications in the Very High Error Channels}

\begin{figure}
\centering
\includegraphics[height=7cm,width=8cm]{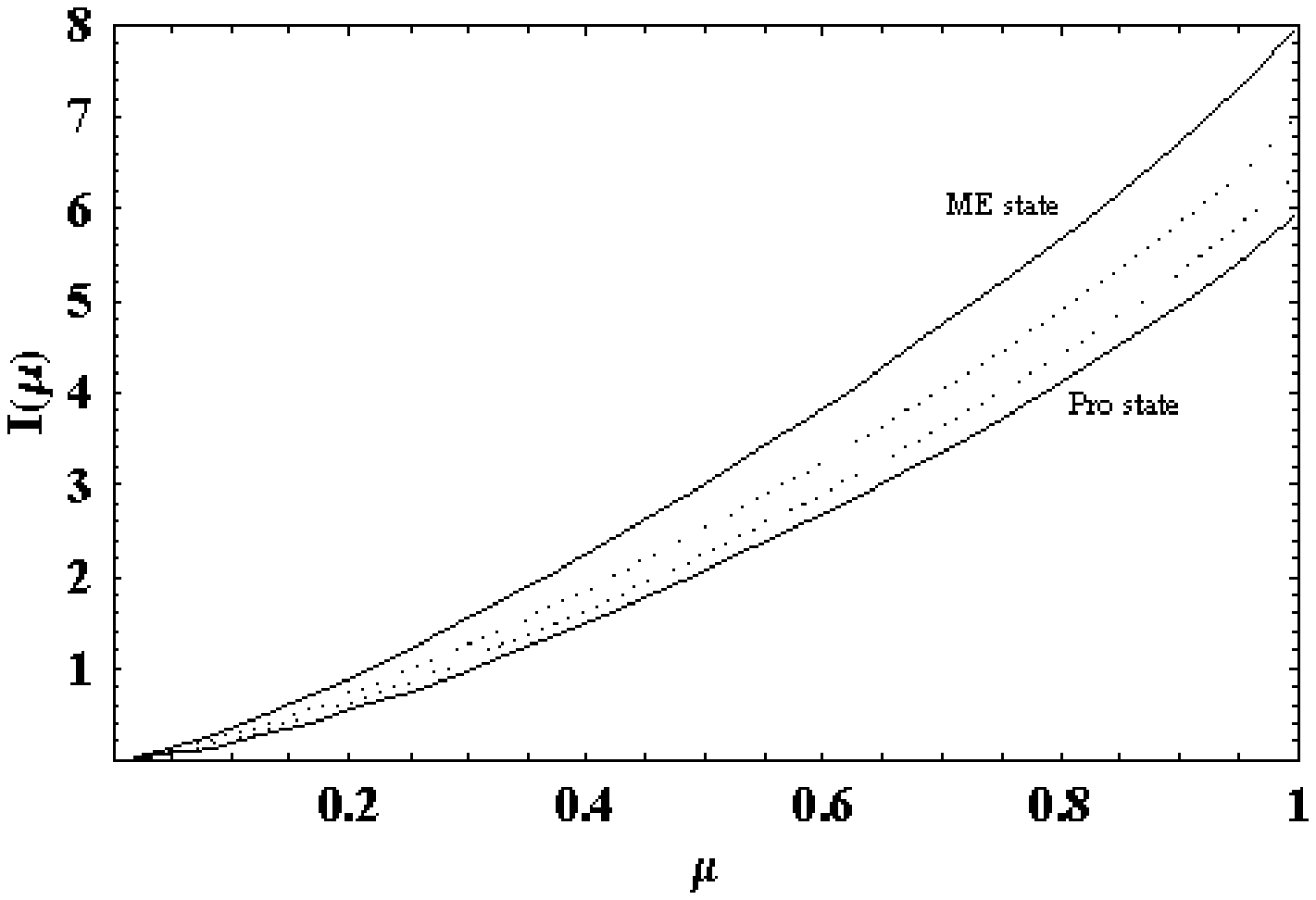}
\caption{ Mutual information $I({\mathcal E}_2(\rho_\theta))$ as a
function of the memory  parameter $\mu$ for very high error
quantum channels $p=q=r=t$ for different values of $k$, in the
$d=4$ dimensional systems. The solid lines represent the mutual
information for maximally entangled ($k=1$) and product states
($k=4$) and the dashed lines are used for non-maximally entangled
states with $k=2, 3$.} \label{Fig.3}
\end{figure}

Here, we would like to compute the maximum amount of information
that can be transmitted through a noisy channel, defined by Eq.
(\ref{Kar}). We shall compare the capacity of the quantum channel
for maximally entangled states (\ref{Maxim}) and the product
states (\ref{Pro}). Mutual information $I(\mathcal E
(\rho_{[l_{i},s]}))$ of the quantum channel for the special cases
$q=r=t$ and $p=q, r=t$ are calculated. Here, we consider very high
error channels for which $p=q=r=t$. Concerning Pauli channel
effects (\ref{Kar}), to optimize the information transmission of
the channel, we must have input states that minimize the output
entropy \cite{Cor,Ban}. For maximally entangled states we have:
\begin{eqnarray*}
&&I^{max-en}(\mathcal E
(\rho_{[l_{i},s]}))=d\log_{2}d\nonumber\\&&+(1-d^{-d})(1-\mu)\times\log_{2}\{(1-\mu)d^{-d}\}
\nonumber\\&&+ \{(1-\mu)d^{-d}+\mu \}\log_{2}\{(1-\mu)d^{-d}+\mu\}
\end{eqnarray*}
This shows that if noises are completely correlated $(\mu=1)$,
then the Holevo limit $\chi^{max-en}(\mathcal E)=Max
I^{max-en}(\mathcal E)=d\log_{2}d$ can be achieved. For the
product states $I^{pro}(\mathcal E)$, we have, in a similar
manner:
\begin{eqnarray*}
&&I^{pro}(\mathcal E)=d\log_{2}d\nonumber\\
&&+ \{(1-\mu)d^{1-d}+\mu
\}\log_{2}\{(1-\mu)d^{-d}+\mu d^{-1}\}\nonumber\\
&&+(1-d^{1-d})(1-\mu)\log_{2}\{(1-\mu)d^{-d}\}.
\end{eqnarray*}
In Figures (\ref{Fig.3}) we compare $I^{max-en}(\mathcal E)$ and
$I^{pro}(\mathcal E)$ schematically, as a function of $\mu$.
These figures show that at high error rates and for channels
without memory $(\mu=0)$, the mutual information is equal to its
minimum values $(I^{max-en}(\mathcal E)=I^{pro}(\mathcal E)=0)$;
on the other hand, in the qudit-environment interaction (no matter
how strong), there exist quantum maximally entangled states, which
are invariant under this interaction, and the mutual information
can attain its maximum value. Hence, for every degree of memory,
these states have better classical information capacity than the
product states. Here, we show that maximally entangled states
(\ref{Maxim}) optimize mutual information transition. In the
channels with high errors, any output density matrix can be
transformed to the following form:
\begin{eqnarray}
\mathcal E (\rho)= (1-\mu)\frac{1}{d^{d}}I^{\otimes d}+\mu \sigma
\end{eqnarray}
(with $Tr\sigma=1, Tr\mathcal E (\rho)=1$). Optimal mutual
information is obtained by minimizing the output entropy, and for
this, we must have a pure state at the output channel.

An indication of the optimality of mutual information and
minimality of entropy is give by $Tr(\rho_{pur})^{2}=1$ in the
output states. Thus, for $\mathcal E (\rho)$ we have:
\begin{eqnarray*}
Tr(\mathcal E (\rho))^{2}= (1-\mu)^{2}\frac{1}{d^{d}}+\mu^{2} Tr
\sigma^{2} + 2\mu(1-\mu)\frac{1}{d^{d}}.
\end{eqnarray*}
The left hand side of the above relation is going to be maximum
for any amount of $\mu$ if $\sigma$ is pure and this happens if
the input state is a maximally entangled state. The optimization
of the Holevo quantity can be achieved by going to an appropriate
bases that diagonalizes $\sigma$. If we assume that $\sigma$ has
$k$ nonzero diagonal elements, then, the entropy is given by:
\begin{eqnarray*}
S(\mathcal E (\rho))&=&k\{(1-\mu)d^{-d}+\mu\frac{1}{k}
\}\log_{2}\{(1-\mu)d^{-d}+\mu\frac{1}{k}\}\nonumber\\
&+&(d^{d}-k)\{(1-\mu)d^{-d}\}\log_{2}\{1-\mu)d^{-d}\}.
\end{eqnarray*}
The minimum value of the above relation can be obtained for $k=1$.
In other words, $\sigma$ must be a pure state, and this happens if
the input state is a maximally entangled state.

\section{Conclusion}
Quantum channels with correlated noise open a new landscape to
quantum communication processes. One of the main applications is
in the standard quantum cryptography BB84 \cite{Benn}, if Alice
and Bob use appropriate states that are suggested in \cite{Jae},
then parties can distillate secure keys (by error correction and
privacy amplification) in the higher amount of quantum bit error
rate (QBER). In other words, $I_{AB}(QBER)>I_{AE}(QBER),
I_{BE}(QBER)$. At this stage QBER is a function of the amount of
memory $QBER=QBER(\mu)$. Hence, this approach can be extended to
quantum key distribution protocols where the key is carried by
quantum states in a space of arbitrary dimension $d$, using two
(or $d+1$) mutually unbiased bases, where for the high memory
channels we have very low error rates. This procedure ensures that
any attempt by any eavesdropper Eve to gain information about the
sender's state induces errors in the transmission, which can be
detected by legitimate parties \cite{Cerf}. In other words, for
arbitrary dimension we can derive the mutual information of
$I_{AB}(\mu),I_{AE}(\mu)$, and $ I_{BE}(\mu)$ and show that in the
higher error channels (no matter how strong), there exist quantum
maximally entangled states, which are invariant under this
interaction. On the other hand, if we are interested in other
quantum key distribution protocols (such as the EPR protocol
\cite{Eke}), we must encode a qubit in a decoherence-free (DF)
subspace of the collective noise for key distribution \cite{Boi1}.
Hence for use of the total dimension of Hilbert space, we must
revise  the EPR protocol for this new approach \cite{Fa}. Another
application of the above extension can be quantum coding, quantum
superdense coding at the higher dimensions, and quantum
teleportation in the Pauli channels with correlated noise.
Although errors in the memory channels can be considered as a
subset of collective noise, which are considered in the DF
approach, some experimental results \cite{Ball} show that in some
special cases the use of these states is appropriate, because in
the memorial channels we make use of all of the maximally
entangled states.

To summarize, we have studied quantum communication channels with
correlated noise in $d$-dimensional systems and have generalized
memory channels for $d$-level systems and have shown that there
exists a memory threshold $\mu_{t}^{f}$, which goes to zero for
high noisy channels. We derived the mutual information of the
quantum channels for maximally entangled states and the product
states for channels with correlated noise. Then, we calculated the
classical capacity of a particular correlated noisy channel and
show that for attaining the Holevo limit we must use $d$ particles
with $d$ degrees of freedom. Furthermore, we chose a special
non-maximally entangled state and showed that in the quantum
depolarizing and quasi-classical depolarizing channels, maximum
classical capacity in the higher memory channels is given by a
maximally entangled state.

\textbf{ACKNOWLEDGMENTS} We would like to thank S. Fallahi for his
helps. This work was supported under project: \emph{ARAM}.

\section{Appendix A: Mutual Information of Maximally Entangled states in the Channel's with Correlated Noise}
We evaluate the action of the channel given by (\ref{Kar}) on the
maximally entangled state (\ref{Maxim}). Before going further, we
would like to define the generalized controlled-\textsc{not} gates
in $d$-dimensional systems. We define the $d$-dimensional unitary
operator
$U_{1,0}=\sigma_{d}=\sum_{k=0}^{d-1}|(k+1)\textnormal{mod}\hspace{0.1cm}d\rangle\langle
k|$ with $\sigma_{d}^{d}=I$ \cite{CM}. This operator should be
calculated to a mod-$d$ adder \cite{Pap}. We will have:
\begin{eqnarray}
\sigma_{d}|j\rangle=|j+1\rangle,\hspace{0.5cm}\textnormal{mod}\hspace{0.2cm}d\nonumber\\
\sigma_{d}^{\dagger}|j\rangle=|j-1\rangle,\hspace{0.5cm}\textnormal{mod}\hspace{0.2cm}d.
\end{eqnarray}
For the case of qubits $(d=2)$, it corresponds to Pauli matrix
$\sigma_{x}=\sigma_{2}$. For every unitary operator $\sigma_{d}$
$(\sigma_{d}^{\dagger})$, the controlled gate $C_{t,l}$
$(C_{t,l}^{\dagger})$, which acts on the site $l$ conditioned on
the site $t$ is naturally defined as:
\begin{eqnarray}
C_{t,l}|i\rangle_{t}\otimes|j\rangle_{l}=|i\rangle_{t}\otimes(\sigma_{d})^{i}|j\rangle_{l}\nonumber\\
C_{t,l}^{\dagger}|i\rangle_{t}\otimes|j\rangle_{l}=|i\rangle_{t}\otimes(\sigma_{d}^{\dagger})^{i}|j\rangle_{l}
\end{eqnarray}
The difference with the qubit case should be taken into account.
In the qubit case the controlled operator just acts when the value
of the site $t$ bit is equal to $1$, whereas in the
$d$-dimensional case, this operator acts $i$ times whenever the
value of the $t$-th qubit is equal to $i$.

For simplicity, we consider $|\psi_{[0,0]}\rangle$ as an input
state and apply some unitary transformation (\textsc{c-not}
operators) on the output density matrix (\ref{Outen}),
\textsc{c-not} operates between first qubit (as the controller)
and the remaining qubits (as the target). We represent
\textsc{c-not} operators as $C_{t,l}$ where the site $t$ is the
controller and the site $l$ is the target. The density matrix
(\ref{Outen}) has two types of sentences, I) the state of first
qudit is $|j_{1}\rangle\langle j_{2}|_{0}$ (include special case
of $j_{1}=j_{2}$) II) the state of first qudit is identity matrix
$I_{0}$.

In the first case, if the density matrix is represented by
$|j_{1}\rangle\langle
j_{2}|_{0}\otimes\sigma_{1}...\otimes\sigma_{i}\otimes...\sigma_{d-1}$,
after \textsc{c-not} operations, the density matrix changes to:
\begin{eqnarray}
&&\bigotimes_{i=1}^{d-1}C_{0,i}^{\dagger}(|j_{1}\rangle\langle
j_{2}|_{0}\otimes\sigma_{1}...\otimes\sigma_{i}\otimes...\sigma_{d-1})\bigotimes_{i=1}^{d-1}C_{0,i}\nonumber\\
&&=|j_{1}\rangle\langle
j_{2}|_{0}\otimes\theta_{1}...\otimes\theta_{i}\otimes...\theta_{d-1}.
\end{eqnarray}
In the above relation, $\bigotimes_{i=1}^{d-1}C_{0,i}$ is the
summary form of $C_{0,1}\otimes C_{0,2}\otimes...\otimes
C_{0,d-1}$ and $\sigma_{i}$ are $|j_{1}\rangle\langle j_{2}|_{i}$
or the identity matrix $I_{i}$ and $\theta_{i}$ are
$|0\rangle\langle 0|_{i}$ or identity matrix $I_{i}$. Under
\textsc{c-not} operations, we have:
\begin{eqnarray}
C_{0,i}^{\dagger}(|j_{1}\rangle\langle
j_{2}|_{0}\otimes|j_{1}\rangle\langle
j_{2}|_{i})C_{0,i}&\longrightarrow&|j_{1}\rangle\langle j_{2}|_{0}\otimes|0\rangle\langle 0|_{i},\nonumber\\
C_{0,i}^{\dagger}(|j_{1}\rangle\langle
j_{2}|_{0}\otimes(\delta_{j_{1}, j_{2}})
I_{i})C_{0,i}&\longrightarrow&|j_{1}\rangle\langle
j_{2}|_{0}\otimes(\delta_{j_{1}, j_{2}}) I_{i}\nonumber\\
&&\textnormal{with}\hspace{0.2cm}i=1,...,d-1.\nonumber\\
\end{eqnarray}
In a similar manner, for the second case of first qudit, which is
in the form
$\sum_{j=0}^{d-1}I_{0}\otimes\sigma_{1}...\otimes\sigma_{i}\otimes...\sigma_{d-1}$
(where $\sigma_{i}$ were previously defined), after \textsc{c-not}
operations, we have:
\begin{eqnarray}
&&\sum_{j=0}^{d-1}\bigotimes_{i=1}^{d-1}C_{0,i}^{\dagger}(I_{0}\otimes\sigma_{1}...\otimes\sigma_{i}\otimes...\sigma_{d-1})\bigotimes_{i=1}^{d-1}C_{0,i}
\nonumber\\&&=\sum_{j=0}^{d-1}I_{0}\otimes\eta_{1}...\otimes\eta_{i}\otimes...\eta_{d-1}
\end{eqnarray}
In the above relation, $\eta_{i}$ are $|j\rangle\langle j|_{i}$ or
the identity matrix $I_{i}$. under \textsc{c-not} operations, we
have:
\begin{eqnarray}
\sum_{j=0}^{d-1}C_{0,i}^{\dagger}(I_{0}\otimes|j\rangle\langle
j|_{i})C_{0,i}&\longrightarrow&\sum_{j=0}^{d-1}I_{0}\otimes|j\rangle\langle
j|_{i},\nonumber\\
\sum_{j=0}^{d-1}C_{0,i}^{\dagger}(I_{0}\otimes
I_{i})C_{0,i}&\longrightarrow&\sum_{j=0}^{d-1}I_{0}\otimes
I_{i},\nonumber\\&&\textnormal{with}\hspace{0.2cm}i=1,...,d-1.\nonumber\\
\end{eqnarray}
The density matrix after \textsc{c-not} operations is given by:
\begin{widetext}
\begin{eqnarray*}
\widetilde{\mathcal
E}^{max-en}(\rho_{[0,0]})&=&\bigotimes_{i=1}^{d-1}C_{0,i}^{\dagger}
{\mathcal
E}^{max-en}(\rho_{[0,0]})\bigotimes_{i=1}^{d-1}C_{0,i}=\frac{(1-\mu)}{d}x\sum_{j_{1},j_{2}=0}^{d-1}|j_{1}\rangle\langle
j_{2}|[(x-y)|0\rangle\langle 0|+yI]^{\otimes(d-1)}\nonumber\\&& +
\frac{(1-\mu)}{d}y\sum_{j_{1},j_{2}=0}^{d-1}|j_{1}\rangle\langle
j_{2}|\sum_{m_{0}=1}^{d-1}[(x-y)\otimes|m_{0}\rangle\langle
m_{0}|+yI]^{\otimes(d-1)}\nonumber\\&& +
\frac{\mu}{d}\sum_{j_{1},j_{2}=0}^{d-1}|j_{1}\rangle\langle
j_{2}|\otimes|0\rangle\langle 0|^{\otimes(d-1)}
-\frac{(1-\mu)}{d}(x-y)I\otimes[(x-y)|0\rangle\langle
0|+yI]^{\otimes(d-1)}\nonumber\\&&
-\frac{(1-\mu)}{d}yI\otimes\sum_{j=0}^{d-1}[(x-y)|j\rangle\langle
j|+yI]^{\otimes(d-1)}\nonumber\\&& +
\frac{(1-\mu)}{d}(x-y+d(q-t))I[(x-y+d(q-t))|0\rangle\langle
0|+(y+dt)I]^{\otimes(d-1)}\nonumber\\&& +
\frac{(1-\mu)}{d}(y+dt)I\otimes\sum_{j=0}^{d-1}[(x-y+d(q-t))|j\rangle\langle
j|+(y+dt)I]^{\otimes(d-1)}.
\end{eqnarray*}
\end{widetext}
with $x$ and $y$ defined as $x=p-q$ and $y=r-t$. After some simple
calculations, the mutual information would be:

\begin{eqnarray}
I^{max-en}(\widetilde{\mathcal E}^{max-en} (\rho_{[l_{i},s]})) &=&
d\log_{2}d+
\sum_{[k_{i}=0]}^{d-1}\Lambda^{0}_{[k_{i}]}\log_{2}[\Lambda^{0}_{[k_{i}]}]\nonumber\\&+&
(d-1)\sum_{[k_{i}=0]}^{d-1}\Lambda^{1}_{[k_{i}]}\log_{2}[\Lambda^{1}_{[k_{i}]}].\nonumber\\
\end{eqnarray}
In the above relation, $[k_{i}]$ represents a set of variables
with $i=1,...,d-1$ and $\Lambda^{l}_{[k_{i}]}$ is given by:
\begin{eqnarray}
\Lambda^{0}_{[k_{i}]}=dB_{[k_{i}]}+A_{[k_{i}]},\hspace{.15cm}
\Lambda^{l}_{[k_{i}]}=A_{[k_{i}]}, \hspace{.15cm}l=1,...,d-1,
\end{eqnarray}

and $A_{[k_{i}]}$ and $B_{[k_{i}]}$ have been defined as:
\begin{widetext}
\begin{eqnarray*}
A_{[k_{i}]}&=&
-\frac{(1-\mu)}{d}(x-y)\prod_{i=1}^{d-1}[(x-y)\delta_{0,k_{i}}+y]\nonumber
-\frac{(1-\mu)}{d}y\sum_{j=0}^{d-1}\prod_{i=1}^{d-1}[(x-y)\delta_{j,k_{i}}+y]\nonumber\\&&
+
\frac{(1-\mu)}{d}(x-y+d(q-t))\prod_{i=1}^{d-1}[(x-y+d(q-t))\delta_{0,k_{i}}+(y+dt)]^{\otimes(d-1)}\nonumber\\&&
+
\frac{(1-\mu)}{d}(y+dt)\sum_{j=0}^{d-1}\prod_{i=1}^{d-1}[(x-y+d(q-t))\delta_{j,k_{i}}+(y+dt)]\nonumber\\
B_{[k_{i}]}&=&\frac{(1-\mu)}{d}x\prod_{i=1}^{d-1}[(x-y)\delta_{0,k_{i}}+y]
+ \frac{(1-\mu)}{d}y\sum_{m_{0}=1}^{d-1}\prod_{i=1}^{d-1}
[(x-y)\delta_{-m_{0},k_{i}}+y] +
\frac{\mu}{d}\prod_{i=1}^{d-1}\delta_{0,k_{i}}.
\end{eqnarray*}
\end{widetext}
\section{Appendix B: Quasi-Classical Depolarizing Channel}
In this appendix, we would like to consider state
$|\Psi_{[0,0]}\rangle=\sum_{j=0}^{d-1} A_{j}|j\rangle^{\otimes
d}$, with $A_{j}\in \mathbb{C}$ as the input state and calculate
the output density matrix and corresponding mutual information for
quasi-classical depolarizing channel with correlated noise.
Similar to the previous appendix, after the calculation of the
output density matrix and applying \textsc{c-not} unitary
transformations on it, the output density matrix is given by:
\begin{widetext}
\begin{eqnarray}\label{B1}
\mathcal E(\rho_{[0,0]})&=& \left\{\mu
a\sum_{j_{1},j_{2}=0}^{d-1}A_{j_{1}}A^{\ast}_{j_{2}}\left[a|j_{1}\rangle\langle
j_{2}|+b\sum_{m=1}^{d-1}|j_{1}+m\rangle\langle
j_{2}+m|\right]\right.\nonumber\\
&&\left.+(1-\mu)(a^{d}-b^{d})\sum_{j=0}^{d-1}|A_{j}|^{2}|j\rangle\langle
j|+(1-\mu)b^{d}\textbf{I}\right\}|0\rangle\langle
0|^{\otimes(d-1)}\nonumber\\
&&+(1-\mu)(a-b)\sum_{j=0}^{d-1}|A_{j}|^{2}|j\rangle\langle
j|\otimes[(a-b)|0\rangle\langle
0|+(1-\mu)b\textbf{I}]^{\otimes(d-1)}\nonumber\\
&&+(1-\mu)b\sum_{l=0}^{d-1}|l\rangle\langle
l|\otimes\sum_{j=0}^{d-1}|A_{j+l}|^{2}[(a-b)|j\rangle\langle
j|+(1-\mu)b\textbf{I}]^{\otimes(d-1)}\nonumber\\
&&-\left\{(1-\mu)(a^{d}-b^{d})\sum_{j=0}^{d-1}|A_{j}|^{2}|j\rangle\langle
j|+(1-\mu)b^{d}\textbf{I}\right\}|0\rangle\langle
0|^{\otimes(d-1)}.
\end{eqnarray}
\end{widetext}
In the above relation $a=pd$ and $b=qd$ and the density matrix has
two types of sentences which are orthogonal to each other. The
matrix elements of the first part (only the first qudit which is
the tensor product to $|0\rangle\langle 0|^{\otimes(d-1)}$) is
given by:
\begin{eqnarray}
E_{k,k}&=&(1-\mu)\left[(a^{d}-b^{d})|A_{k}|^{2}+b^{d}\right]\nonumber\\&+&\mu\left[(a-b)|A_{k}|^{2}+b\right],
\nonumber\\
E_{k,k'}&=&\mu a A_{k}A^{\ast}_{k'}+\mu
b\sum_{m=1}^{d-1}A_{k-m}A^{\ast}_{k'-m}.
\end{eqnarray}
The explicit form of the matrix elements for the state suggested
in the Eq. (\ref{SUST}) would be:
\begin{eqnarray}
t=E_{0,0}&=&
(1-\mu)\left[(a^{d}-b^{d})\cos^{2}\alpha+b^{d}\right]\nonumber\\&+&\mu\left[(a-b)\cos^{2}\alpha+b\right],
\nonumber\\
s=E_{0,k}&=&E_{k,0}=\mu a
\frac{\cos\alpha\sin\alpha}{\sqrt{d-1}}\nonumber\\&+&\mu
b\left[\frac{\sin^{2}\alpha}{d-1}(d-2)+\frac{\cos\alpha\sin\alpha}{\sqrt{d-1}}
\right]\hspace{0.5cm}k \geq1,\nonumber\\
c=E_{k,k}&=&(1-\mu)\left[(a^{d}-b^{d})\frac{\sin^{2}\alpha}{d-1}+b^{d}\right]\nonumber\\&+&\mu\left[(a-b)\frac{\sin^{2}\alpha}{d-1}+b\right]
\hspace{0.5cm}k \geq1,\nonumber\\
r=E_{k,k'}&=&\mu a \frac{\sin^{2}\alpha}{d-1}+\mu
b\left[\frac{\sin^{2}\alpha}{d-1}(d-3)+2\frac{\cos\alpha\sin\alpha}{\sqrt{d-1}}
\right]\nonumber\\&&
\hspace{0.5cm}k\neq k' \geq1.\nonumber\\
\end{eqnarray}
The second part of the density matrix is diagonal and is given by:
\begin{widetext}
\begin{eqnarray}
A_{[k_{i}]}&=&(1-\mu)(a-b)\left[\cos^{2}\alpha\delta_{0,k_{0}}+\frac{\sin^{2}\alpha}{d-1}(1-\delta_{0,k_{0}})\right]
\prod_{i=1}^{d-1}[(a-b)\delta_{0,k_{i}}+b]\nonumber\\&&
+(1-\mu)b\sum_{j=0}^{d-1}\left[\cos^{2}\alpha(\delta_{0,k_{0}+j}+\delta_{d,k_{0}+j})+\frac{\sin^{2}\alpha}{d-1}
[1-(\delta_{0,k_{0}+j}+\delta_{d,k_{0}+j})]\right]
\times\prod_{i=1}^{d-1}\left[(a-b)\delta_{j,k_{i}}+b\right]\nonumber\\&&
-\left\{(1-\mu)(a^{d}-b^{d})\left[\cos^{2}\alpha\delta_{0,k_{0}}+\frac{\sin^{2}\alpha}{d-1}(1-\delta_{0,k_{0}})\right]
+(1-\mu)b^{d}\right\}\prod_{i=1}^{d-1}\delta_{0,k_{i}}.
\end{eqnarray}
\end{widetext}
We apply the above density matrix to the calculation of the mutual
information, as a function of $\alpha,\mu,d,a,$ and $b$. It is
given by:
\begin{eqnarray*}
&&I(\mathcal E, \alpha,\mu,d,a,b)= d\log_{2}d+
(d-2)\lambda^{0}\log_{2}\lambda^{0}\nonumber\\&&+
\lambda^{1}\log_{2}\lambda^{1}+\lambda^{2}\log_{2}\lambda^{2}
+\sum_{[k_{i}=0]}^{d-1}A_{[k_{i}]}\log_{2}A_{[k_{i}]}
\end{eqnarray*}
In the above relation $\lambda^{0}$ [with the degeneracy of
$(d-2)$] and $\lambda^{1,2}$ are the eigenvalues of the first part
of the density matrix and are given by:
\begin{eqnarray*}
\lambda^{0}&=&c-r,
\nonumber\\\lambda^{1,2}&=&\frac{1}{2}\left\{t+c+(d-2)r
\pm\left[t^{2}+4(d-1)s\right.\right.\nonumber\\
&&-\left.\left.2tc+c^{2}-2(d-2)(c-t)r+(d-2)^{2}r^{2}\right]^{1/2}\right\}
\end{eqnarray*}
In the following, we consider
$|\Psi_{[0,0]}\rangle=\sum_{j=0}^{d-1} A_{j}|j\rangle^{\otimes d}$
with coefficients that are suggested in Eq. (\ref{KM1}) for the
case $d=4$, as the input state, and use the output density matrix
(\ref{B1}), then the mutual information as a function of
$\theta,\mu,d,a,$ and $b$, is given by:
\begin{eqnarray}\label{MUKM}
&&I(\mathcal E, \theta,\mu,d,a,b)= d\log_{2}d+
2\lambda^{0}\log_{2}\lambda^{0}\\&&+
\lambda^{1}\log_{2}\lambda^{1}+\lambda^{2}\log_{2}\lambda^{2}
+\sum_{[k_{i}=0]}^{d-1}D_{[k_{i}]}\log_{2}[D_{[k_{i}]}]\nonumber
\end{eqnarray}
In the above relation $\lambda^{0}$ (with the degeneracy of $2$)
and $\lambda^{1,2}$ are eigenvalues of the first part of the
density matrix and $D_{[k_{i}]}$ is given by:
\begin{widetext}
\begin{eqnarray}\label{D}
D_{[k_{i}]}&=&\frac{(1-\mu)}{4}(a-b)\left[(1+3\cos^{2}\theta)\delta_{0,k_{0}}+\sin^{2}\theta(1-\delta_{0,k_{0}})\right]
\prod_{i=1}^{d-1}[(a-b)\delta_{0,k_{i}}+b]\nonumber\\&&
+\frac{(1-\mu)}{4}b\sum_{j=0}^{d-1}\left[(1+3\cos^{2}\theta)(\delta_{0,k_{0}+j}+\delta_{d,k_{0}+j})+\sin^{2}\theta
[1-(\delta_{0,k_{0}+j}+\delta_{d,k_{0}+j})]\right]
\times\prod_{i=1}^{d-1}\left[(a-b)\delta_{j,k_{i}}+b\right]\nonumber\\&&
-\left\{\frac{(1-\mu)}{4}(a^{d}-b^{d})\left[(1+3\cos^{2}\theta)\delta_{0,k_{0}}+\sin^{2}\theta(1-\delta_{0,k_{0}})\right]
+(1-\mu)b^{d}\right\}\prod_{i=1}^{d-1}\delta_{0,k_{i}},
\end{eqnarray}
\end{widetext}
\begin{eqnarray}\label{lamda}
\lambda^{0}&=&-w+z+f, \nonumber\\\lambda^{1,2}&=&w+2z+f
\pm\left[3v^{2}+4w^{2}-2wz+z^{2}\right]^{1/2},
\end{eqnarray}
\begin{eqnarray}\label{wv}
w&=&\frac{\mu}{4}\sin^{2}\theta,\hspace{0.5cm}v=\mu\frac{(a-b)}{2}\cos\theta\sin\theta\nonumber\\z&=&\left\{\frac{\mu(a-b)}{4}+(1-\mu)\frac{(a^{4}-b^{4})}{4}\right\}\cos^{2}\theta\nonumber\\
f&=&\mu\left[\frac{(a-b)}{4}+b\right]+(1-\mu)\left[\frac{(a^{4}-b^{4})}{4}+b^{4}\right].
\end{eqnarray}

\section{Appendix C: Depolarizing Channel}
Similar to Appendix B, the output density matrix and the
corresponding mutual information for the depolarizing channel with
correlated noise, is given by:
\begin{widetext}
\begin{eqnarray*}
\mathcal E(\rho_{[0,0]})&=& \left\{\left[(1-\mu)(x-y)^{d}+\mu
x\right]\sum_{j_{1},j_{2}=0}^{d-1}A_{j_{1}}A^{\ast}_{j_{2}}|j_{1}\rangle\langle
j_{2}|+\mu
y\sum_{j_{1},j_{2}=0}^{d-1}\sum_{m=1}^{d-1}A_{j_{1}-m}A^{\ast}_{j_{2}-m}
|j_{1}\rangle\langle
j_{2}|\right.\nonumber\\
&&\left.+(1-\mu)(x^{d}-y^{d}-(x-y)^{d})\sum_{j=0}^{d-1}|A_{j}|^{2}|j\rangle\langle
j|+(1-\mu)y^{d}\textbf{I}\right\}|0\rangle\langle
0|^{\otimes(d-1)}\nonumber\\
&&+(1-\mu)(x-y)\sum_{j=0}^{d-1}|A_{j}|^{2}|j\rangle\langle
j|\otimes[(x-y)|0\rangle\langle
0|+y\textbf{I}]^{\otimes(d-1)}\nonumber\\
&&+(1-\mu)y\sum_{l=0}^{d-1}|l\rangle\langle
l|\otimes\sum_{j=0}^{d-1}|A_{j+l}|^{2}[(x-y)|j\rangle\langle
j|+y\textbf{I}]^{\otimes(d-1)}\nonumber\\
&&-\left\{(1-\mu)(x^{d}-y^{d})\sum_{j=0}^{d-1}|A_{j}|^{2}|j\rangle\langle
j|+(1-\mu)y^{d}\textbf{I}\right\}|0\rangle\langle
0|^{\otimes(d-1)}
\end{eqnarray*}
\end{widetext}
In the above relation $x=p+(d-1)q$ and $y=qd$ and the density
matrix has two types of sentences which are orthogonal to each
other. The matrix elements of the first part (only the first
qudit, which is the tensor product to $|0\rangle\langle
0|^{\otimes(d-1)}$) is given by:
\begin{eqnarray}\label{B2}
E_{k,k}&=&(1-\mu)\left[(x^{d}-y^{d})|A_{k}|^{2}+y^{d}\right]\nonumber\\&&+\mu\left[(x-y)|A_{k}|^{2}+y\right],
\nonumber\\
E_{k,k'}&=&\left[(1-\mu)(x-y)^{d}\right.\nonumber\\&&\left.+\mu
x\right] A_{k}A^{\ast}_{k'}+\mu
y\sum_{m=1}^{d-1}A_{k-m}A^{\ast}_{k'-m}.
\end{eqnarray}
The explicit form of the matrix elements for the state that are
suggested in Eq. (\ref{SUST}) would be:
\begin{eqnarray}
t=E_{0,0}&=&[(1-\mu)(x-y)+\mu x]\cos^{2}\alpha+\mu
y\sin^{2}\alpha\nonumber\\
&&+\left[(1-\mu)(x^{d}-y^{d}-(x-y)^{d})\right]\cos^{2}\alpha\nonumber\\
&&+(1-\mu)y^{d},\nonumber\\
s=E_{0,k}&=&E_{k,0}=[(1-\mu)(x-y)^{d}+\mu
x]\frac{\cos\alpha\sin\alpha}{\sqrt{d-1}}\nonumber\\
&&+\mu
y\left[\frac{\sin^{2}\alpha}{d-1}(d-2)+\frac{\cos\alpha\sin\alpha}{\sqrt{d-1}}
\right],\nonumber\\&&\hspace{4cm}k \geq1,\nonumber\\
c=E_{k,k}&=&[(1-\mu)(x-y)^{d}+\mu x]\frac{\sin^{2}\alpha}{d-1}\nonumber\\
&&+\mu y\left[\frac{\sin^{2}\alpha}{d-1}(d-2)+\cos^{2}\alpha
\right]\nonumber\\&&+\left[(1-\mu)(x^{d}-y^{d}-(x-y)^{d})\right]\frac{\sin^{2}\alpha}{d-1}\nonumber\\
&&+(1-\mu)y^{d}
\hspace{0.5cm}k \geq1,\nonumber\\
r=E_{k,k'}&=&[(1-\mu)(x-y)^{d}+\mu
x]\frac{\sin^{2}\alpha}{d-1}\nonumber\\
&&+\mu
y\left[\frac{\sin^{2}\alpha}{d-1}(d-3)+2\frac{\cos\alpha\sin\alpha}{\sqrt{d-1}}
\right]\nonumber\\
&&\hspace{4cm}k\neq k' \geq1.
\end{eqnarray}
The second part of the density matrix is diagonal and is given by:
\begin{widetext}
\begin{eqnarray}
A_{[k_{i}]}&=&(1-\mu)(x-y)\left[\cos^{2}\alpha\delta_{0,k_{i}}+\frac{\sin^{2}\alpha}{d-1}(1-\delta_{0,k_{i}})
\right]\prod_{i=1}^{d-1}[(x-y)\delta_{0,k_{i}}+y]\nonumber\\&&+
(1-\mu)y\sum_{j=0}^{d-1}\left[\cos^{2}\alpha(\delta_{0,k_{0}+j}+\delta_{d,k_{0}+j})+\frac{\sin^{2}\alpha}{d-1}
[1-(\delta_{0,k_{0}+j}+\delta_{d,k_{0}+j})]\prod_{i=1}^{d-1}[(x-y)\delta_{j,k_{i}}+y]\right]\nonumber\\&&
-(1-\mu)\left\{(x^{d}-y^{d})\cos^{2}\alpha\delta_{0,k_{i}}+\frac{\sin^{2}\alpha}{d-1}(1-\delta_{0,k_{i}})
+(1-\mu)y^{d} \right\}\prod_{i=1}^{d-1}\delta_{0,k_{i}}.
\end{eqnarray}
\end{widetext}
We apply the above density matrix to calculate the mutual
information as a function of $\alpha,\mu,d,x,$ and $y$. It is
given by:
\begin{eqnarray*}
&&I(\mathcal E, \alpha,\mu,d,a,b) = d\log_{2}d+
(d-2)\lambda^{0}\log_{2}\lambda^{0}\nonumber\\&&+
\lambda^{1}\log_{2}\lambda^{1}+\lambda^{2}\log_{2}\lambda^{2}
+\sum_{[k_{i}=0]}^{d-1}A_{[k_{i}]}\log_{2}A_{[k_{i}]}.
\end{eqnarray*}
In the above relation $\lambda^{0}$ (with the degeneracy of
$(d-2)$) and $\lambda^{1,2}$ are eigenvalues of the first part of
the density matrix which are given by:
\begin{eqnarray*}
\lambda^{0}&=&c-r,
\nonumber\\\lambda^{1,2}&=&\frac{1}{2}\left\{t+c+(d-2)r
\pm\left[t^{2}+4(d-1)s\right.\right.\nonumber\\
&&-\left.\left.2tc+c^{2}-2(d-2)(c-t)r+(d-2)^{2}r^{2}\right]^{1/2}\right\}.
\end{eqnarray*}

Similar to the previous section, we consider
$|\Psi_{[0,0]}\rangle=\sum_{j=0}^{d-1} A_{j}|j\rangle^{\otimes d}$
with coefficients that are suggested in Eq. (\ref{KM1}) for the
case $d=4$, as the input state and calculate the mutual
information for the quantum depolarizing channel as a function of
$\theta,\mu,d,x,$ and $y$ for states that are suggested in
(\ref{KM1}). The mutual information would be given by Eq.
(\ref{MUKM}), by replacing $a\longrightarrow x$ and
$b\longrightarrow y$ and a similar expression for $D_{[k_{i}]}$
(\ref{D}) and $\lambda^{0,1,2}$ (\ref{lamda}), with the following
variables for $w,v,z$, and $f$:
\begin{eqnarray*}
w&=&\frac{1}{4}\left[(1-\mu)(x^{4}-y^{4})+\mu\right]\sin^{2}\theta,\nonumber\\
v&=&\frac{1}{2}\left[(1-\mu)(x^{4}-y^{4})+\mu(x-y)\right]\cos\theta\sin\theta
\nonumber\\
z&=&\left\{\frac{\mu(x-y)}{4}+(1-\mu)\frac{(x^{4}-y^{4})}{4}\right\}\cos^{2}\theta\nonumber\\
f&=&\mu\left[\frac{(x-y)}{4}+y\right]+(1-\mu)\left[\frac{(x^{4}-y^{4})}{4}+y^{4}\right].
\end{eqnarray*}



\begin{references}
\bibitem {Fa1} In another paper, we have derived an exact expression
for the transition of $d$-level systems through channels with
correlated noise, using Gottesman unitary operators.
\bibitem {Com} For product states, we have some different choice, if we consider
the computational basis ${|s\rangle}$ (with $s= 0,..., d-1$) and
its dual under a Fourier transformation:
$|\psi^{p}_{[l_{i},s]}\rangle= |s\rangle
\frac{1}{d^{\frac{d-1}{2}}}\sum_{[k_{i}]}^{d-1} e^{ \frac{2 \pi
i}{d}\sum_{i=1}^{d-1}k_{i}l_{i}}|k_{1}\rangle...|k_{d-1}\rangle $,
as input states to quantum channel, where
$|\psi^{p}_{[l_{i},s]}\rangle$ is one set of $(d^{d})$ orthonormal
product states of d qudits, the fidelity for the product states is
given by: $F^{p}_{[l_{i},s]} = (1-\mu)[p+ (d-1)q]^{d} + \mu p $,
which is not an efficient computational basis. Hence, have a lower
mutual information in comparison with the computational basis in
the $\mu=0$ case.
\bibitem {KMCom} It is not
hard to show that their approach is not so exact \cite{KM}. The
derivations in the \cite{KM} indicate that all the states have the
same values of entanglement which interpolate continuously between
the product basis and the maximally entangled basis by varying the
parameters $\theta_{\alpha}$. They have also found explicitly such
bases in three, four and five dimensional spaces and have given a
general solution for the spaces with arbitrary dimensions. But as
can be seen in four dimensional case, we should revise their
coefficients to achieve the Holevo limit. Hence, for example, if
we consider their coefficients for arbitrary $d$-level systems as
input state of our channel, for all values of $\theta_{\alpha}$,
the mutual information doesn't get the Holevo limit, whereas, the
maximally entangled states that were suggested in (\ref{Maxim})
achieve the Holevo limit for $\mu=1$.
\end{references}
\end{document}